\documentclass[12pt]{article}
\usepackage{graphicx}
\usepackage{amsfonts}
\usepackage{amssymb}
\newcommand{\nin}{\noindent}
\newcommand{\be}{\begin{equation}}
\newcommand{\ee}{\end{equation}}
\newcommand{\bea}{\begin{eqnarray}}
\newcommand{\eea}{\end{eqnarray}}

\newcommand{\hf}{\frac{1}{2}}
\newcommand{\nn}{\nonumber\\}

\newcommand{\ol}{\overline}

\begin{document}
\begin{flushleft}
KCL-PH-TH/2010-21
\end{flushleft}
\vspace{0.5cm}
\begin{center}

{\bf{\Large Emergent relativistic-like Kinematics and Dynamical Mass Generation for a Lifshitz-type Yukawa model}}

\vspace{1cm}

J.~Alexandre, N.~E.~ Mavromatos and D.~Yawitch\\
\emph{King's College London, Department of Physics, Theoretical Physics, London WC2R 2LS, UK}

\vspace{2cm}

{\bf Abstract}

\end{center}

\vspace{0.5cm}

\nin We study the Infra Red (IR) limit of dispersion relations for scalar and fermion fields in a Lifshitz-type Yukawa model,
after dressing by quantum fluctuations. Relativistic-like dispersion relations emerge dynamically in the IR regime of the model,
after quantum corrections are taken into account.
In this regime, Dynamical mass generation also takes place, but in such a way that the particle excitations remain
massive, even if the bare masses vanish. 
The group velocities of the corresponding massive particles of course are smaller than the speed of light,
in a way consistent with the IR regime where the analysis is performed. We also comment on possible extensions
of the model where the fermions are coupled to an Abelian gauge field.

\vspace{2cm}

\section{Introduction}

Quantum field theories in the Lifshitz context exhibit interesting renormalization properties \cite{field}.
Lifshitz type models are based on an anisotropy between space and time directions, which is characterized by the
dynamical critical exponent z, determining the properties of space-time coordinates under scale transformations
$t \to b^z t$ and ${\bf x} \to b {\bf x}~.$
For $z > 1$ the higher powers of momentum in the propagators lower the superficial degree of divergence of graphs,
yielding the renormalizability of new interactions, such as the four-fermion interaction \cite{4Fermi}. In the specific case of
$z=3$ in 3+1 dimensions, the scalar field is dimensionless, so that any function of the field is renormalizable.
 The case of an exponential potential for the scalar field has been studied in \cite{LL}, where the exact effective potential
has been derived and found consistent with the
well-known Liouville theory in 1+1 dimensions.
Moreover, in this framework, divergences of renormalizable interactions in the Standard Model become softer,
for instance in the Yukawa model \cite{Yukawa} only logarithmic divergences appear. Finally, we mention that this approach has also been proposed as a renormalizable alternative to Einstein General Relativity~\cite{horava}, which could lead to a consistent model of Quantum Gravity.

While absent from the classical action, Lorentz symmetry can naturally be generated in  Lifshitz-type models through
quantum corrections, since the corresponding kinetic term is a relevant operator and dominates the dispersion
relation of the modes in the infrared regime (IR).
In \cite{iengo}, the authors study the case $z=2$, for scalar theories involving derivative interactions. There,
logarithmic divergences imply a speed of light running with a mass scale,
and the discrepancy in the IR speeds of light which are obtained for different species of particles is discussed.

In the current work we shall study a softer version of the Lifshitz framework, namely
a $z=3$ Lifshitz-type Yukawa model, where the only divergence appears in the scalar mass.
As a consequence, there is no running speed of light in the model, while the pertinent IR dispersion relations are
consistent between different species of particles.
However, an important feature of the model, which will be discussed in section 2, is the fact that
the quadratic term in the scalar-field derivatives has the ``wrong sign'', so that an apparent tachyonic mode
develops in this theory. Nevertheless, as a consequence of the six-order derivative term, which has the correct sign,
the tachyonic mode disappears and a relativistic
dispersion relation can be obtained after expanding the frequency around its minimum. This is similar to an expansion
around the Fermi surface in condensed matter physics, and leads to a momentum shift.

The structure of the article is as follows: in section 2, we discuss the fate of a
tachyonic mode to sixth order in derivatives of the scalar field, and the recovery of a relativistic-like dispersion relation
for the relevant excitations in the model. The corresponding effective light cone depends on the particle species, but we 
explain that no contradiction with the speed of light arises. 
In section 3, we study the non-perturbatively dressed masses of the system, and show that these do not vanish, even if the bare masses
go to zero. We show this by means of a non-perturbative treatment of the model, in the form
of a differential Schwinger-Dyson approach, based on a set of
self consistent coupled equations for the dressed parameters. We exhibit
numerical solutions for these self consistent equations, where one can see that the no IR divergence occurs, unlike
the case of the perturbative analysis based on a massless bare theory.

In section 4, we discuss an extension of the model where the fermions couple to an Abelian
Gauge field (a form of anisotropic Quantum Electrodynamics - $QED$).
We show that the one-loop analysis is not sufficient to allow for an interpretation of the gauge boson as a photon
in such a model, in the sense that its group velocity is different from that of light \emph{in vacuo}.
Conclusions and Outlook are presented in section 5, where we stress that the dynamical mass generation in our
Yukawa-Lifshitz model allows fermions to have a mass, without Higgs mechanism,
and that the possibility of formulating a Lifshitz Standard Model may lead to phenomenologically relevant realizations.

\section{IR dispersion relations}

We explain in this section the different features related to the IR of the
dressed theory, starting from a Lifshitz-type Yukawa model. For the sake of clarity,
the corresponding calculations can be found in Appendix A.

\subsection{The Model}

The Lifshitz-type Yukawa model in $D+1$ dimensions and with anisotropic scaling $z$ is defined by~\footnote{Throughout this work we use units where the speed of light \emph{in vacuo} is $c=1$.}
\bea
S_{D,z}&=&\int dt d^Dx\Big\lbrace \frac{1}{2}(\dot\phi)^2-\frac{1}{2}(\vec\partial\phi)\cdot(\Delta^{z-1}\vec\partial)\phi
+ i\ol\psi\gamma^0\dot\psi-i\ol\psi\Delta^\frac{z-1}{2}(\vec\gamma\cdot\vec\partial)\psi\nn
&&~~~~~~~~~~~~~~~~~-\frac{1}{2}m_0^{2z}\phi^2-m_0^z\ol\psi\psi-g_0\phi\ol\psi\psi\Big\rbrace ,
\eea
where $\Delta\equiv \partial^i\partial^j\delta_{ij}$, with $i,j$ spatial indices.
We consider here the case $z=D=3$, where the mass dimensions of coordinates and fields are
\be
[t]=-3~~~~~~[x]=-1~~~~~~[\phi]=0~~~~~~~[\psi]=\frac{3}{2}.
\ee
The bare action is then
\bea\label{bare}
S&=&\int dt d^3x\Big\lbrace \frac{1}{2}(\dot\phi)^2-\frac{1}{2}(\vec\partial\phi)\cdot(\Delta^2\vec\partial)\phi
+ i\ol\psi\gamma^0\dot\psi-i\ol\psi\Delta(\vec\gamma\cdot\vec\partial)\psi\nn
&&~~~~~~~~~~~~~~~~~-\frac{1}{2}m_0^6\phi^2-m_0^3\ol\psi\psi-g_0\phi\ol\psi\psi\Big\rbrace ,
\eea
where the Yukawa coupling $g_0$ has mass dimension $[g_0]=3$ and $m_0$ is the bare mass of particles.
We consider the same bare mass for the scalar and the fermions: a different mass would involve a new parameter
which would not change qualitatively the non-perturbative analysis we present in the next section.
This theory is super-renormalizable, and the only divergence occurs in the dressed scalar mass \cite{Yukawa}.

\subsection{Naive approach}

Quantum fluctuations generate lower orders in space derivatives of the field,
such that the effective theory is
\bea
S_{eff}&=&\int dt d^3x\Bigg\lbrace \frac{1+\zeta_\phi}{2}(\dot\phi)^2
-\frac{\lambda_\phi}{2}(\vec\partial\phi)\cdot(\vec\partial\phi)-\frac{M^6}{2}\phi^2\Bigg\}\\
&&+\int dt d^3x\Bigg\lbrace i(1+\zeta_\psi)\ol\psi\gamma^0\dot\psi-i\lambda_\psi\ol\psi(\vec\gamma\cdot\vec\partial)\psi
-m^3\ol\psi\psi\Bigg\rbrace+S_{int}\nonumber
\eea
where $M$ and $m$ are respectively the dressed scalar and fermion masses, and $\lambda_\phi,\lambda_\psi,\zeta_\phi,\zeta_\psi$
arise from quantum corrections. $S_{int}$ contains interactions, higher orders derivative terms, derivative interactions and
higher orders in fermion fields.\\
If one considers the free bosonic and fermionic sectors individually, then there are several possible ways to rescale space time coordinates
and fields, in order to recover the relativistic dispersion relations. But because of the interactions contained in $S_{int}$,
the space time coordinates have to be rescaled in the same way for the bosonic and the fermionic sectors. Indeed, the Yukawa interaction,
as well as other interactions generated by quantum fluctuations, are local and occur at the same event in space time.
As a consequence, we define the global rescaling
\be\label{rescaling}
t\to at~~~~{\bf x}\to b{\bf x}~~~~\phi\to A\phi~~~~\psi\to B\psi,
\ee
where the mass dimensions are $[b]=[B]=0$, $[a]=-2$ and $[A]=-1$, in such a way that the new mass dimensions of
space-time coordinates and fields
are the expected ones in an isotropic theory. The new quadratic terms of the effective action, then, become:
\bea
&&ab^3\int dtd^3{\bf x}\Bigg\{A^2\left( \frac{1+\zeta_\phi}{2a^2}(\dot\phi)^2
-\frac{\lambda_\phi}{2b^2}(\vec\partial\phi)\cdot(\vec\partial\phi)
-\frac{M^6}{2}\phi^2\right) \\
&&~~~~~~~~~~~~~~~+B^2\left( i\frac{1+\zeta_\psi}{a}\ol\psi\partial_0\gamma^0\psi
-i\frac{\lambda_\psi}{b}\ol\psi(\vec\gamma\cdot\vec\partial)\psi
-m^3\ol\psi\psi\right) \Bigg\}\nonumber
\eea
It is easy to check that one cannot set all the coefficients of the derivative terms to unity in a consistent way.
Nevertheless, one can impose the requirement that the coefficients of the time derivatives be equal to unity, which leads to the following constraints:
\be\label{constraints}
(1+\zeta_\phi)A^2b^3=a~~~~\mbox{and}~~~~(1+\zeta_\psi)B^2b^3=1~.
\ee
We find that no further constraint can be imposed in a consistent way.
We are then led to the following quadratic terms in the effective action
\bea\label{quadratic}
&&\int dt d^3{\bf x}\Bigg\{\frac{1}{2}(\dot\phi)^2-\frac{v_\phi^2}{2}(\vec\partial\phi)\cdot(\vec\partial\phi)-\frac{\tilde M^2}{2}\phi^2\nn
&&~~~~~~~~~~~~~~~~+i\ol\psi\partial_0\gamma^0\psi-iv_\psi\ol\psi(\vec\gamma\cdot\vec\partial)\psi-\tilde m\ol\psi\psi\Bigg\},
\eea
where we defined
\bea\label{define}
v_\phi^2=\lambda_\phi A^2ab~~~~&,&~~~~v_\psi=\lambda_\psi B^2ab^2\nn
\tilde M^2=M^6A^2ab^3~~~~&,&~~~~\tilde m=m^3 B^2ab^3.
\eea
Note that the constraints (\ref{constraints}) do not fix the parameters $a,b,A,B$ in a unique way. The new parameters (\ref{define})
(which are not independent)
are renormalized parameters, fixed in principle by the ``experiment'', and define a specific rescaling, with unique values for $a,b,A,B$.

\subsection{Non-trivial scalar dispersion relation}

The previous arguments hold if $\lambda_\phi>0$. However, as demonstrated in Appendix A, the one-loop
expressions for $\lambda_\phi$ and $\lambda_\psi$ in the model are
\be\label{lambdasoneloop}
\lambda_\phi^{(1)}=\frac{-g_0^2I}{\pi^2m_0^2}~<~0~~~~\mbox{and}~~~~\lambda_\psi^{(1)}=\frac{3g_0^2J}{16\pi^2m_0^4}
\ee
with
\bea
I&=&\int_0^\infty \frac{dx~x^6}{(1+x^6)^{3/2}}\left(\frac{8}{3}-\frac{3(3x^6+7)}{4(1+x^6)}+\frac{15x^6}{2(1+x^6)^2}\right)
\simeq 0.25\nn
J&=&\int_0^\infty \frac{x^{10}dx}{(1+x^6)^{5/2}}\simeq 0.16
\eea
As a consequence, the scalar dispersion relation is of the form (before rescaling fields and space time coordinates)
\be
(1+\zeta_\phi)\omega^2=M^6-|\lambda_\phi| k^2+\eta_\phi k^4+(1+\xi_\phi)k^6+{\cal O}(k^8),
\ee
where $\eta_\phi$ and $\xi_\phi$ are quantum corrections, such that
\be
\eta_\phi\propto g_0^2/m_0^4~~~~\mbox{and}~~~~\xi_\phi\propto g_0^2/m_0^6.
\ee
This dispersion relation presents an apparent tachyonic sign, but
because of the positive term $k^6$, the frequency has actually a non-trivial minimum for some
momentum ${\bf k}_0$ such that
\be\label{k_0^2}
k^2_0={\bf k}_0\cdot{\bf k}_0=\sqrt{\frac{|\lambda^{(1)}_\phi|}{3}}+{\cal O}(g_0^2/m_0^4).
\ee
One can therefore expand $\omega^2$ in powers of~\footnote{One can recall at this stage a similar situation
that characterizes condensed-matter systems, linearized about their Fermi surfaces.} ${\bf q}={\bf k}-{\bf k}_0$
\be
\omega^2\simeq\omega_0^2+\frac{1}{2}\left.\frac{d^2\omega^2}{dk^2}\right|_{{\bf k_0}}q^2,
\ee
and  we obtain from eq.(\ref{k_0^2})
\bea
\omega^2&\simeq&\mu^6+4|\lambda_\phi^{(1)}|q^2+{\cal O}(g_0/m_0)^3\\
\mbox{with}~~~~\mu^6&=&M^6-3\left(\frac{|\lambda_\phi^{(1)}|}{3}\right)^{3/2}=M^6+{\cal O}(g_0/m_0)^3\nonumber,
\eea
such that this momentum shift implies the replacement of $\lambda_\phi^{(1)}$ by $4|\lambda_\phi^{(1)}|$ in the 
new dispersion relation, based on the shifted momentum ${\bf q}$.
We check now that this shifted momentum cannot be detected {\it at one loop} (order $g_0^2$). 
Indeed, the corresponding kinetic term in the action is, in Fourier components,
\bea\label{shift}
&&\frac{1}{2}\int\frac{d\omega}{2\pi}\frac{d^3{\bf k}}{(2\pi)^3}~
\Big(\omega^2 -4|\lambda_\phi^{(1)}|({\bf k}-{\bf k}_0)^2-\mu^6\Big)\phi_{\omega,{\bf k}}\phi_{-\omega,-{\bf k}} \\
&=&\frac{1}{2}\int\frac{d\omega}{2\pi}\frac{d^3{\bf k}}{(2\pi)^3}~
\Big(\omega^2 -4|\lambda_\phi^{(1)}| k^2-\mu^6\Big)\phi_{\omega,{\bf k}+{\bf k}_0}\phi_{-\omega,-{\bf k}-{\bf k}_0}\nn
&=&\frac{1}{2}\int\frac{d\omega}{2\pi}\frac{d^3{\bf k}}{(2\pi)^3}~
\Big(\omega^2 -4|\lambda_\phi^{(1)}| k^2-M^6\Big)\phi_{\omega,{\bf k}}\phi_{-\omega,-{\bf k}}+{\cal O}(g_0/m_0)^3,\nonumber
\eea
since $\lambda_\phi={\cal O}(g_0^2/m_0^2)$ and $k_0^2={\cal O}(g_0/m_0)$, and the linear term in ${\bf k}_0$ is a surface term.
From eq.(\ref{shift}), the scalar dispersion relation is, {\it at one loop},
\be
\omega^2\simeq M^6+4|\lambda_\phi^{(1)}| k^2
\ee
We finally note that the fermionic dispersion relation always involves $\lambda_\psi^2$, and is therefore independent of the sign
of $\lambda_\psi$.

\subsection{Group velocities and effective Yukawa coupling}

From the quadratic action (\ref{quadratic}), we can obtain the IR dispersion relations for the fermionic or bosonic excitations in the model
\be\label{disprel}
\omega^2=\mu^2+v^2k^2+{\cal O}(k/\mu)^2,
\ee
where $v$ is a generic speed ($v_\psi$ or $v_\phi$) and $\mu$ is a generic {\it dressed} mass ($\tilde m$ or $\tilde M$).
Note that the speed $v$ {\it cannot} be identified with the speed of light. Indeed, the speed of light would be obtained
from the dispersion relation (\ref{disprel}) in the limit where $\mu\to0$, but we will show in the next section that the scalar and fermion
dressed masses never vanish, even if the bare mass $m_0$ goes to zero, as a consequence of dynamical mass generation. Instead, one can find
from the relation (\ref{disprel}) the product of the group velocity $v_g$ and the phase velocity $v_p$ of particles:
\be\label{vgvp}
v_gv_p=\frac{d\omega}{dk}~\frac{\omega}{k}=v^2+{\cal O}(k/\mu)^2~.
\ee
A word of caution is in order at this point. One should keep in mind that this dispersion relation is valid for $k<<\mu$ only.
In this regime of momenta, $v$ cannot represent a limiting speed of particles
in the model, since the latter can only be obtained in the limit $k\to\infty$.\\
From the constraints (\ref{constraints}) and the definitions (\ref{define}), we find the following relation between $v_\phi$ and $v_\psi$
\be\label{vpsivphi}
v_\psi=v_\phi~\frac{\lambda_\psi}{1+\zeta_\psi}\sqrt\frac{1+\zeta_\phi}{4|\lambda_\phi|}~.
\ee
Using the one-loop expressions (\ref{lambdasoneloop}), we can readily see that the speeds $v_\psi$ and $v_\phi$ satisfy the relation
\be\label{ineq1}
\frac{v_\psi}{v_\phi}\simeq\frac{\lambda_\psi^{(1)}}{2\sqrt{|\lambda_\phi^{(1)}|}}\simeq\frac{3J}{32\pi\sqrt I}~\frac{g_0}{m_0^3}<<1
\ee
Finally, one can express the dressed dimensionless Yukawa coupling $\tilde g$ in terms of the different parameters
obtained after the rescaling (\ref{rescaling}). From the interaction
\be
\tilde g~\phi\ol\psi\psi\equiv g~ab^3AB^2~\phi\ol\psi\psi,
\ee
and the relations (\ref{constraints},\ref{define}), we obtain
\be
\tilde g=g~\frac{v_\phi^{3/2}(1+\zeta_\phi)^{1/4}}{|4\lambda_\phi|^{3/4}(1+\zeta_\psi)}.
\ee
We show in Appendix A that the one loop coupling $g^{(1)}$ is equal to the bare coupling $g_0$, so that
the regime remains perturbative at one loop ($\tilde g<<1$), if $g_0v_\phi^{3/2}<< |4\lambda_\phi^{(1)}|^{3/4}$ or, equivalently:
\be\label{ineq2}
v_\phi<<\sqrt{\frac{4I}{\pi^2}} ~\frac{g_0^{1/3}}{m_0}\simeq 0.31~\frac{g_0^{1/3}}{m_0}
\ee
It is interesting to observe (\ref{ineq1}) that the speed $v_\psi$ of the fermions is smaller than the corresponding one, $v_\phi$, for the scalar fields. However, the reader should recall (c.f. (\ref{vgvp})) that $v_\psi^2$ and $v_\phi^2$ represent the respective products of the group and the phase velocities and thus should not be confused with the propagation velocity of the physical excitations. Nevertheless, in the infrared, the corresponding group velocities
are: 
\be
v_g^2 =  \frac{v^2}{\sqrt{\frac{\mu^2}{k^2} + v^2}} \simeq v^2\frac{k}{\mu}
\ee
since $\mu^2/k^2 \gg v^2$. As we shall see in the next session, the effective masses $\mu^2$ of fermions and scalars are of the same order, and thus, for a given momentum scale, on account of (\ref{ineq1}), the group and phase velocities of the fermions will be smaller than the corresponding ones of the scalars. Moreover, the inequalities (\ref{ineq1}) and (\ref{ineq2}) show that the model can represent ``slowly'' moving particles only, in the
perturbative regime $g_0^{1/3}/m_0<<1$. \\
Finally, the relativistic-like dispersion relation (\ref{disprel}) exhibits a different effective light cone for the scalar and the fermion 
(although we stress again that the speed of light is not questioned by our analysis), showing that this model can be relevant to Condensed Matter,
as we will discuss in the conclusion.

\section{Non-perturbatively dressed masses}

In this section we shall demonstrate that the dressed masses of the model cannot vanish, even if the bare masses do. This dynamical generation
can be studied only through a non-perturbative approach (such as the Schwinger-Dyson approach), since the resulting
masses are not analytic functions of the coupling constant, and no perturbative expansion can exhibit the dynamical
mass-generation mechanism.\\
We will quantize the theory (\ref{bare}), and derive an exact self-consistent equation for the proper graphs generator
functional $\Gamma$,
in the form of a differential Schwinger-Dyson equation, giving the evolution of $\Gamma$ with the amplitude of the bare mass $m_0$.
As explained in the original paper \cite{original}, this approach consists in controlling the amplitude of quantum fluctuations:
if one starts from a large mass $m_0$, quantum fluctuations are ``frozen'', and the system almost remains classical, $\Gamma\simeq S$.
As the bare parameter $m_0$ decreases, quantum fluctuations gradually dress the system, and it can be shown~\cite{original} that
the corresponding flows in $m_0$ are equivalent, at one-loop, to those given by the standard Callan-Symanzik equations.\\
We first review the path integral quantization of the model, in order to define our notations, and then derive the
exact evolution equation for the proper graph generating functional $\Gamma$ with the bare mass $m_0$. Then, upon assuming a specific
functional form for $\Gamma$, we derive the corresponding dressed scalar and fermion masses.
The pertinent evolution equations are non-perturbative. They constitute a set of self consistent coupled equations, thereby
representing a resummation of graphs.

\subsection{Path integral quantization}

The quantum theory is based on the partition function
\bea
Z[j,\eta,\ol\eta]&=&\int{\cal D}[\phi,\ol\psi,\psi]\exp\left( iS+i\int dtd^3x (j\phi+\ol\eta\psi+\ol\psi\eta)\right)\nn
&=&\exp(iW[j,\eta,\ol\eta]),
\eea
where $j,\eta,\ol\eta$ are the sources, and $W$ is the connected graphs generator functional.
The classical fields are defined as
\be\label{ident1}
\phi_{cl}=\frac{-i}{Z}\frac{\delta Z}{\delta j},~~~~~~
\psi_{cl}=\frac{-i}{Z}\frac{\delta Z}{\delta\ol\eta},~~~~~~
\ol\psi_{cl}=\frac{i}{Z}\frac{\delta Z}{\delta\eta},
\ee
and we have
\bea
\frac{\delta^2W}{\delta j\delta j}&=&i\phi_{cl}\phi_{cl}-i<\phi\phi>\nn
\frac{\delta^2W}{\delta\eta\delta\ol\eta}&=&i\ol\psi_{cl}\psi_{cl}-i<\ol\psi\psi>,
\eea
where
\be
<\cdots>\equiv\frac{1}{Z}\int{\cal D}[\phi,\ol\psi,\psi](\cdots)
\exp\left( iS+i\int dtd^3x (j\phi+\ol\eta\psi+\ol\psi\eta)\right)
\ee
The proper graphs generator functional $\Gamma[\phi_{cl},\psi_{cl},\ol\psi_{cl}]$ is defined as the
Legendre transform of $W$
\be
\Gamma=W-\int dtd^3x (j\phi_{cl}+\ol\eta\psi_{cl}+\ol\psi_{cl}\eta),
\ee
where the sources must be viewed as functionals of the classical fields, and therefore functions of the bare parameters in the model.
From this definition, we obtain
\be
\frac{\delta\Gamma}{\delta\phi_{cl}}=-j,~~~~~~
\frac{\delta\Gamma}{\delta\psi_{cl}}=\ol\eta,~~~~~~
\frac{\delta\Gamma}{\delta\ol\psi_{cl}}=-\eta,
\ee
and
\be\label{ident2}
(\delta^2\Gamma)_{\phi_{cl}\phi_{cl}}=-(\delta^2W)^{-1}_{jj},~~~~~~
(\delta^2\Gamma)_{\psi_{cl}\ol\psi_{cl}}=-(\delta^2W)^{-1}_{\eta\ol\eta},
\ee
where the notation $(\delta^2A)_{ij}$ represents the $(i,j)$-element of the matrix with components equal to the second derivatives of $A$.

\subsection{Exact evolution equation}

From these definitions and properties, we can now derive the exact evolution equation of $\Gamma$ with $m_0$. In
what follows, we denote a derivative with respect to $m_0$ with a dot.
The first step is to note that
\bea
\dot\Gamma&=&\dot W+\int dt d^3x\left( \frac{\delta W}{\delta j}\partial_{m_0}j+\frac{\delta W}{\delta\eta}\dot\eta
+\dot{\ol\eta}\frac{\delta W}{\delta\ol\eta}\right)\nn
&&-\int dt d^3x\left(\partial_{m_0}j\phi_{cl}+\dot{\ol\eta}\psi_{cl}+\ol\psi_{cl}\dot\eta\right) =\dot W
\eea
Using the identities (\ref{ident1}) to (\ref{ident2}), then, we obtain
\bea\label{evolG}
\dot\Gamma&=&-3m_0^2\int dt d^3x\left( m_0^3<\phi^2>+<\ol\psi\psi> \right)\nn
&=& -3m_0^2\int dt d^3x\left(m_0^3\phi_{cl}^2+\ol\psi_{cl}\psi_{cl} \right)\nn
&&+3im_0^2\mbox{Tr}\left\lbrace m_0^3 (\delta^2\Gamma)^{-1}_{\phi_{cl}\phi_{cl}}
+(\delta^2\Gamma)^{-1}_{\psi_{cl}\ol\psi_{cl}}  \right\rbrace.
\eea
We stress that the self consistent evolution equation (\ref{evolG}) for $\Gamma$ is exact, and not
based on any assumption or expansion. The resummation of all the quantum corrections to the bare action is contained in the trace.
Note that, if one replaces $\Gamma$ by the bare action $S$ in this trace, we obtain the usual expression for the one-loop theory,
after integration over $m_0$.

\subsection{Ansatz for the proper graph generating functional}

From now on, we omit the subscript $cl$ on the classical fields.
In order to get physical insight on the quantum theory obtained from the solution of eq.(\ref{evolG}),
we need to assume a functional form for $\Gamma$. We consider the following ansatz
\bea\label{ansatz}
\Gamma&=&\int dt d^3x\Bigg\lbrace \frac{1}{2}(\dot\phi)^2-\frac{1}{2}(\vec\partial\phi)\cdot(\Delta^2\vec\partial\phi)
+i\ol\psi\gamma^0\dot\psi-i\ol\psi\Delta(\vec\gamma\cdot\vec\partial)\psi\nn
&&~~~~~~~~~~~~~~~~-V(\phi)-U(\phi)\ol\psi\psi\Bigg\rbrace,
\eea
where $U(\phi), V(\phi)$ are scalar potentials to be determined. The ansatz (\ref{ansatz}) ignores lower order kinetic terms
generated by quantum fluctuations, since these are negligible compared to $k^6$ in the UV regime relevant for the loop
integral (trace in the evolution equation (\ref{evolG})). Also, in the framework of the gradient expansion,
we neglect higher order derivatives, which are also generated by quantum corrections. As a consequence, we
concentrate on the scalar potential sector (self coupling potential $V(\phi)$), and the scalar-fermion coupling $U(\phi)\ol\psi\psi$.

We show in Appendix B that, after plugging the ansatz (\ref{ansatz}) into the exact evolution equation
(\ref{evolG}), the evolution equations for the potentials are
\bea\label{dotUV}
\dot V&=&3m_0^5\phi^2+\frac{m_0^5}{\pi^2}\ln\left( \frac{2\Lambda^3}{\sqrt{V''}}\right)
+\frac{m_0^2U}{\pi^2}\ln\left( \frac{2\Lambda^3}{U}\right)\nn
\dot U&=&3m_0^2+\frac{m_0^2}{8\pi^2}\frac{[U']^2}{(V''-U^2)^2}\ln\left(\frac{V''}{U^2}\right)(V''+U^2-2m_0^3U)\\
&&-\frac{m_0^2}{8\pi^2}\frac{m_0^3U''+2[U']^2}{V''-U^2}+\frac{m_0^5}{8\pi^2}\frac{U(UU''+2[U']^2)}{V''(V''-U^2)}\nonumber
\eea
where a prime denotes a derivative with respect to the scalar field.
Note that the apparent singularities in the evolution of $U$, when $U^2\to V''$, actually cancel each other.
We indeed checked that, after setting $V''=U^2(1+\epsilon)$, we obtain
\be
\dot U=3m_0^2-\frac{m_0^5[U']^2}{8\pi^2U^3} +{\cal O}(\epsilon),
\ee
so that the evolution equation is well defined.

\subsection{Polynomial expansion of the potentials}

We consider the following expansion for the potentials
\bea\label{expUV}
V(\phi)&=&V_0+\kappa\phi+\frac{M^6}{2}\phi^2+{\cal O}(\phi^3)\nn
U(\phi)&=&m^3+g\phi+{\cal O}(\phi^2),
\eea
where $m,M$ are respectively the fermion and scalar dressed masses,
and $g$ is the dressed Yukawa coupling.
The linear term $\kappa\phi$ is purely generated by quantum fluctuations, and arises from the trace
in eq.(\ref{evolG}) which contains $(\delta^2\Gamma)^{-1}_{\ol\psi\psi}$.
The field-independent term $V_0$
takes care of the divergences present in the evolution equation for $V(\phi)$, since setting $\phi=0$ in the latter equation gives
\be
\dot V_0=\frac{m_0^5}{\pi^2}\ln\left( \frac{2\Lambda^3}{M^3}\right)
+\frac{m_0^2}{\pi^2}m^3\ln\left( \frac{2\Lambda^3}{m^3}\right)~,
\ee
and the evolution equations for the remaining parameters $g,m,M$, obtained by differentiation of eqs.(\ref{dotUV}),
become then divergence-free. Only the evolution equation for $\kappa$ still contains a divergence:
\be
\dot\kappa=\frac{gm_0^2}{\pi^2}\left[ \ln\left( \frac{2\Lambda^3}{m^3}\right) -1\right]~.
\ee
However, since this parameter does not
enter into the other evolution equations for $g,m,M$, the latter have cut-off independent flows in $m_0$.
The parameter $\kappa$ decouples from the rest and does not appear in any loop calculation, since the linear term in
$\phi$ is not present in the bare theory.\\
Plugging the expansions (\ref{expUV}) into the evolutions equations
(\ref{dotUV}), we obtain a set of non-linear coupled differential equations for $g,m,M$:
\bea\label{finaleq}
M^5\dot M&=&m_0^5-\frac{g^2}{6\pi^2}\frac{m_0^2}{m^3}\\
m^2\dot m&=&m_0^2+\frac{g^2m_0^2}{4\pi^2}\frac{M^6+m^6-2m_0^3m^3}{(M^6-m^6)^2}\ln\left( \frac{M}{m}\right)\nn
&&+\frac{g^2}{12\pi^2}\frac{m_0^2}{M^6-m^6}\left(\frac{m_0^3m^3}{M^6}-1\right) \nn
\dot g&=&\frac{3g^3m_0^2}{2\pi^2}\left[ \frac{3m^3-m_0^3}{(M^6-m^6)^2}+\frac{4m^6(m^3-m_0^3)}{(M^6-m^6)^3}\right]
\ln\left( \frac{M}{m}\right) \nn
&&+\frac{g^3}{4\pi^2}\frac{m_0^2}{(M^6-m^6)^2}\left( m_0^3\frac{m^6}{M^6}-\frac{M^6}{m^3}+3(m_0^3-m^3)\right)~. \nonumber
\eea
Once again, one can check that the apparent singularities when $M^6\to m^6$ actually do not occur: if one sets $M^6=m^6(1+\epsilon)$,
an expansion in $\epsilon$ shows that the singularities cancel out in the limit $\epsilon\to 0$.

As expected,
one can also check that the one-loop results (calculated
in Appendix A) are recovered after integration over $m_0$, if one replaces the dressed parameters in
the right-hand side of the evolution equations (\ref{finaleq}) by the bare ones:
\bea\label{oneloop}
M^6_{(1)}&=&m_0^6+\frac{g_0^2}{\pi^2}\left[ \ln\left( \frac{\Lambda}{m_0}\right) +\frac{\ln2-1}{3}\right]\nn
m^3_{(1)}&=&m_0^3+\frac{g_0^2}{24\pi^2m_0^3}\nn
g^{(1)}&=&g_0~,
\eea
where we note that the one-loop correction to the coupling vanishes.

\nin In figs.~1 and 2 we plot the numerical solutions of eqs.(\ref{finaleq}) in the range of bare masses from $m_0/\Lambda=1$ down to $m_0=0$, for
$g_0/\Lambda^3=0.01$. We observe that the dressed masses do \emph{not} vanish when the bare mass $m_0$ goes to zero, due to dynamical mass generation. This is an important physical feature of the model.

\begin{figure}[!htbp]
\begin{center}
\includegraphics[width=70mm]{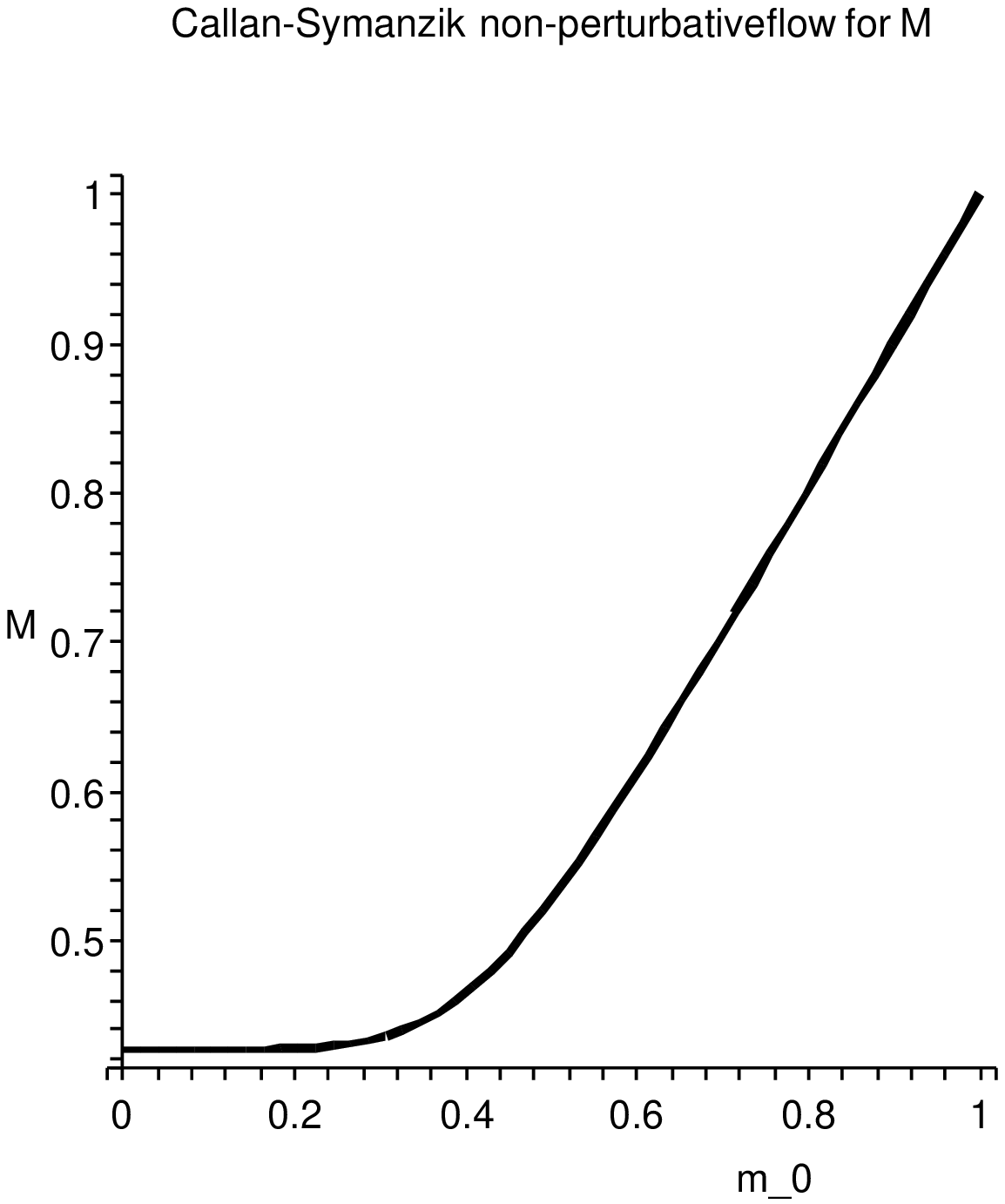}\includegraphics[width=70mm]{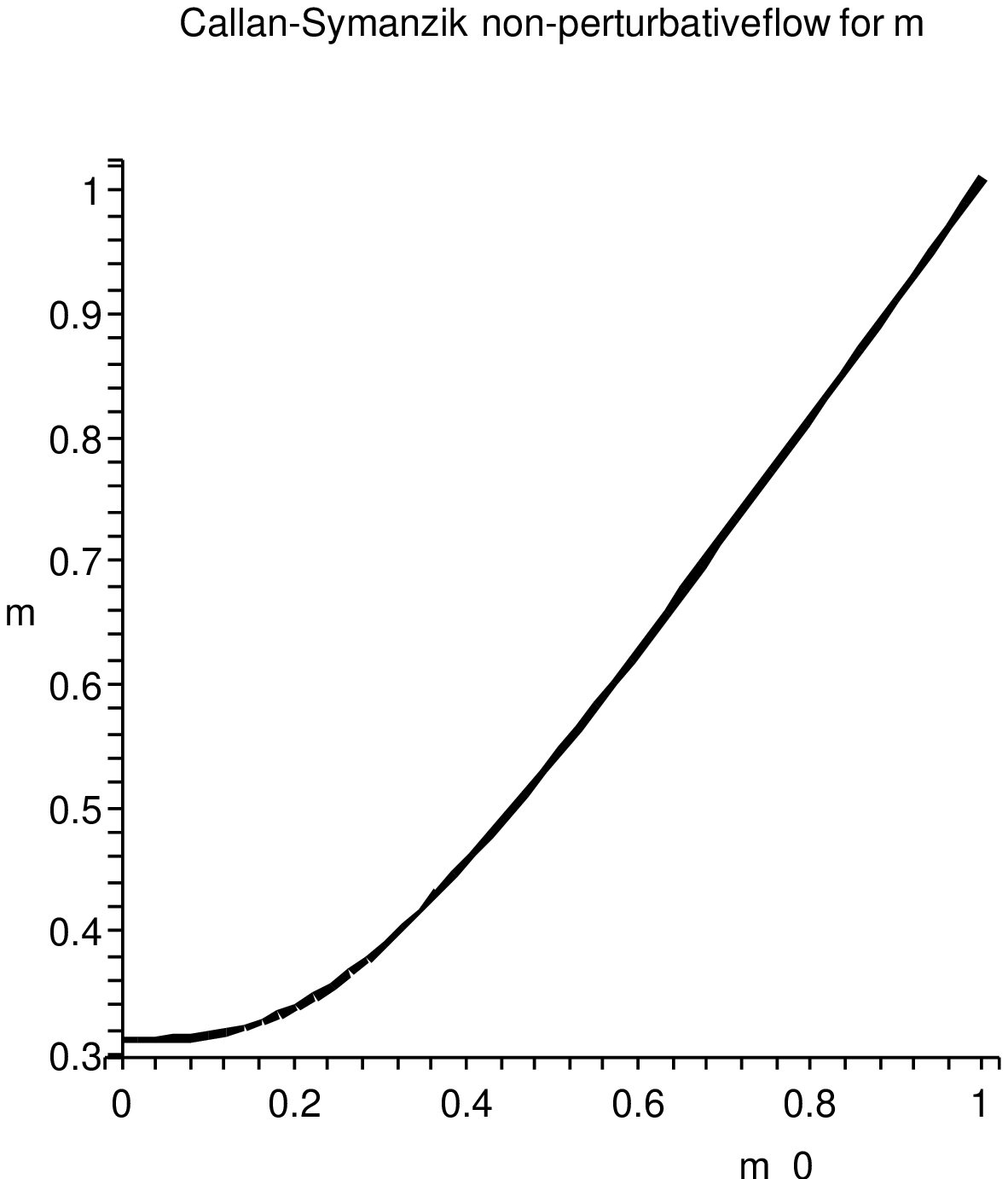}
\end{center}
\caption{Non-perturbative flow for the dressed masses $M$ (scalar), $m$ (fermion) versus $m_0$.
The dressed masses do not vanish when the bare mass $m_0$ goes to zero, due to dynamical mass generation.}
\label{Cs_flow1}
\end{figure}

\begin{figure}[!htbp]
\begin{center}
\includegraphics[width=70mm]{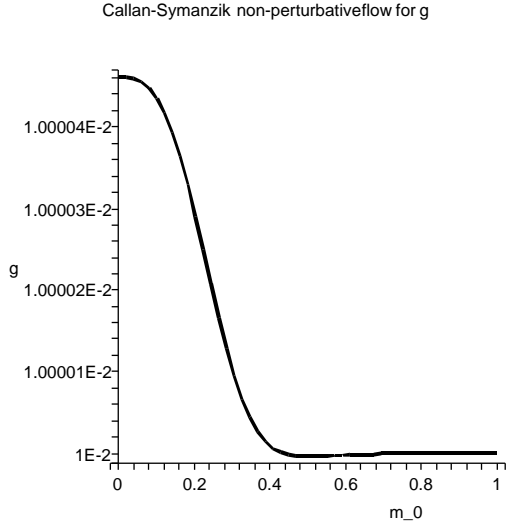}
\end{center}
\caption{Non-perturbative flow for the dressed coupling $g$ versus $m_0$.
Note that the one-loop correction to the coupling vanishes, explaining the
small change in values for $g$.}
\label{Cs_flow2}
\end{figure}

\subsection{IR stability}

One can see from the one-loop expressions (\ref{oneloop}) that the limit $m_0\to 0$ leads to IR divergences. This is actually the
case at any order of the loop expansion. Our numerical analysis shows that the resummation provided by the set of the coupled equations
(\ref{finaleq}) restores the IR stability of the system, and that no divergence occurs in the dressed system, when $m_0\to 0$.
Figs.~3 and 4 compare the one-loop and non-perturbative flows of $M$ and $m$ in the range of bare masses from  $m_0/\Lambda=1$ down to $m_0=0$
(the region around $m_0=0$ is zoomed in, to highlight the difference).
We clearly see with the fermion mass that the would-be IR divergence, in the limit where $m_0\to0$, does not
occur in the non-perturbative flow.
This feature is a consequence of dynamical mass generation, which was already studied in the context of (2+1)-dimensional quantum electrodynamics ($QED_3$), following the present
non-perturbative method \cite{QED3}.

\begin{figure}[!htbp]
\begin{center}
\includegraphics[width=70mm]{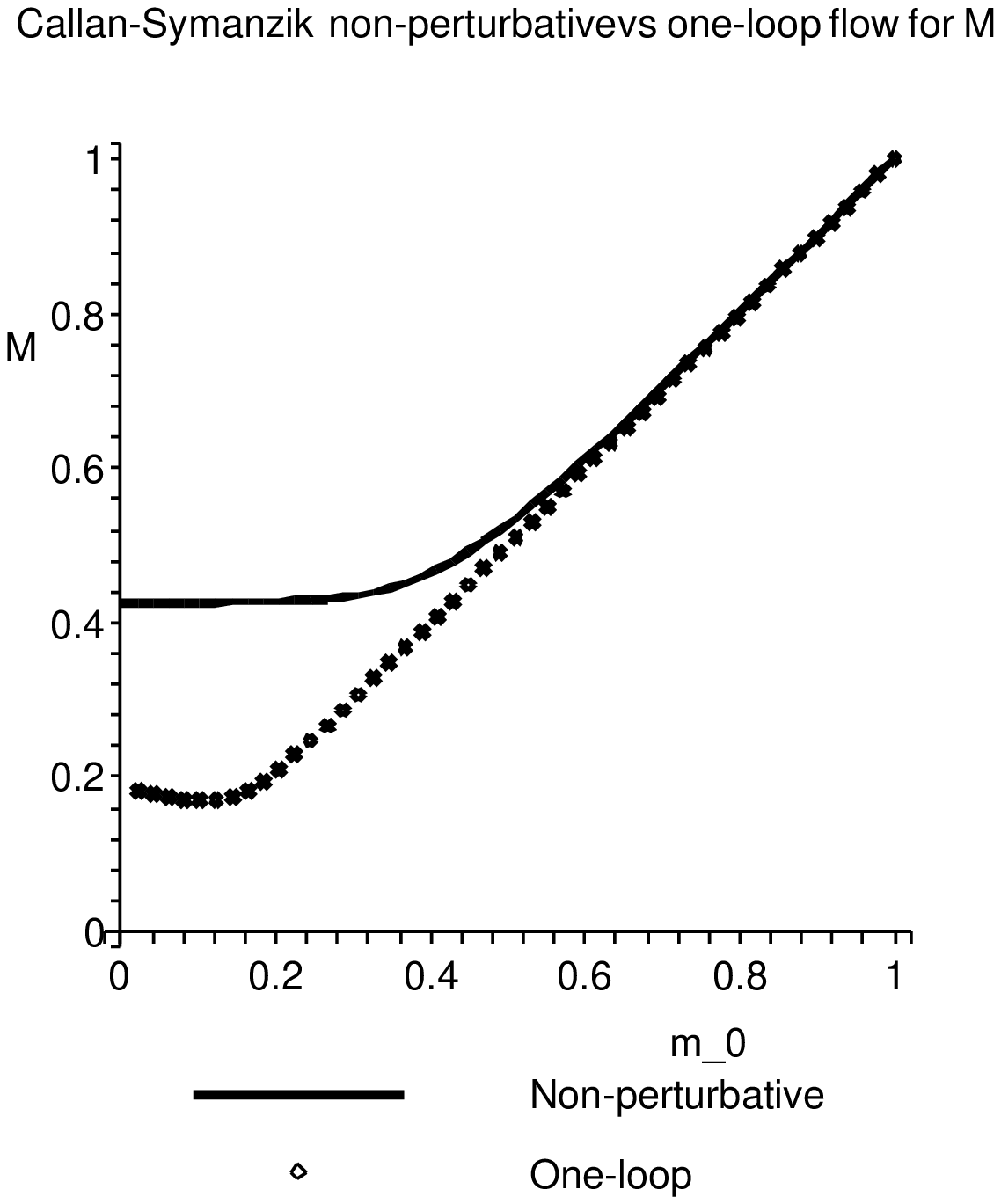}\includegraphics[width=70mm]{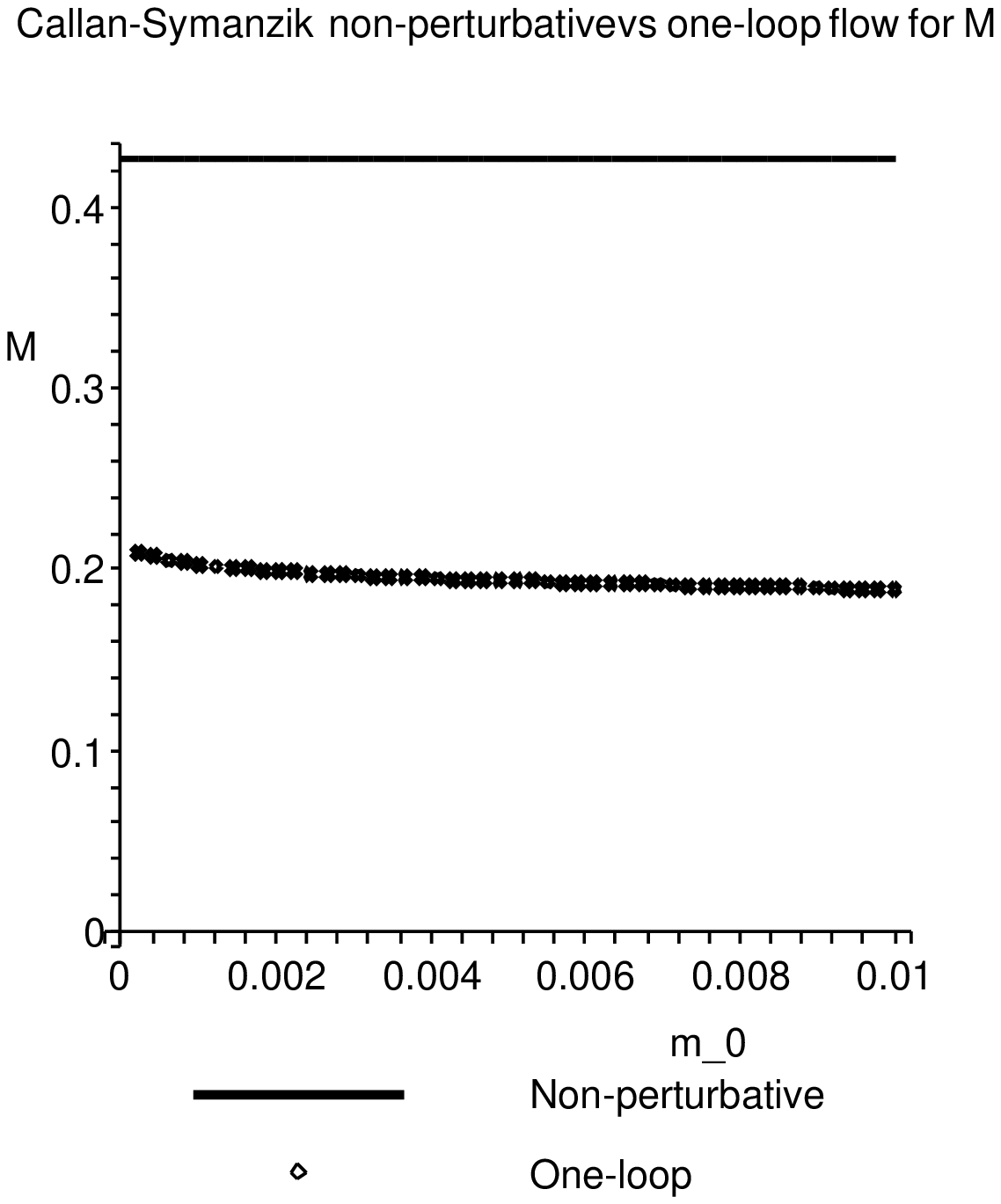}
\end{center}
\caption{Comparison between the one-loop and the non-perturbative flows for the scalar dressed
mass $M$ versus $m_0$ (the figure on the right is a zoom near $m_0=0$)}
\label{M_comp}
\end{figure}

\begin{figure}[!htbp]
\begin{center}
\includegraphics[width=70mm]{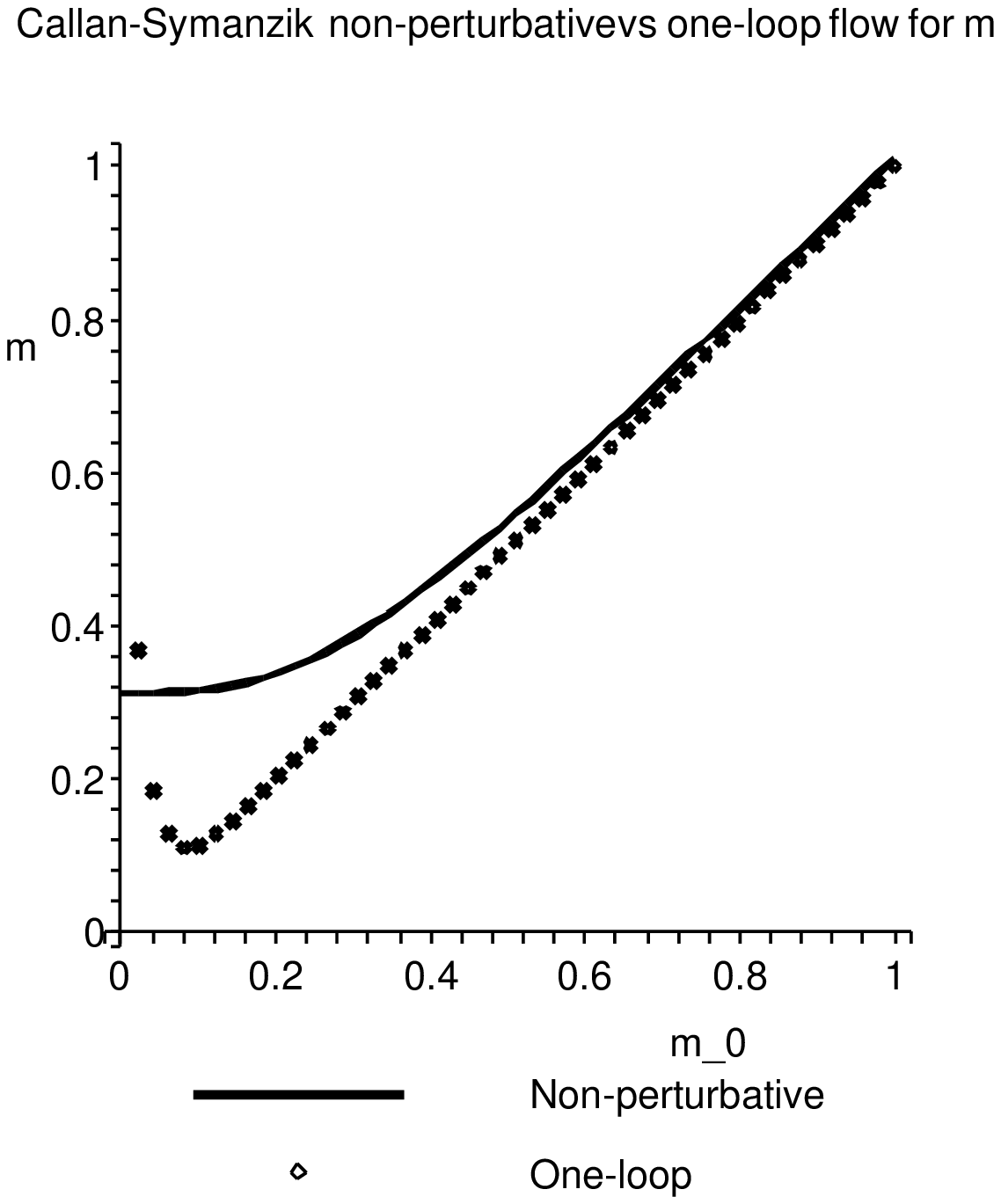}\includegraphics[width=70mm]{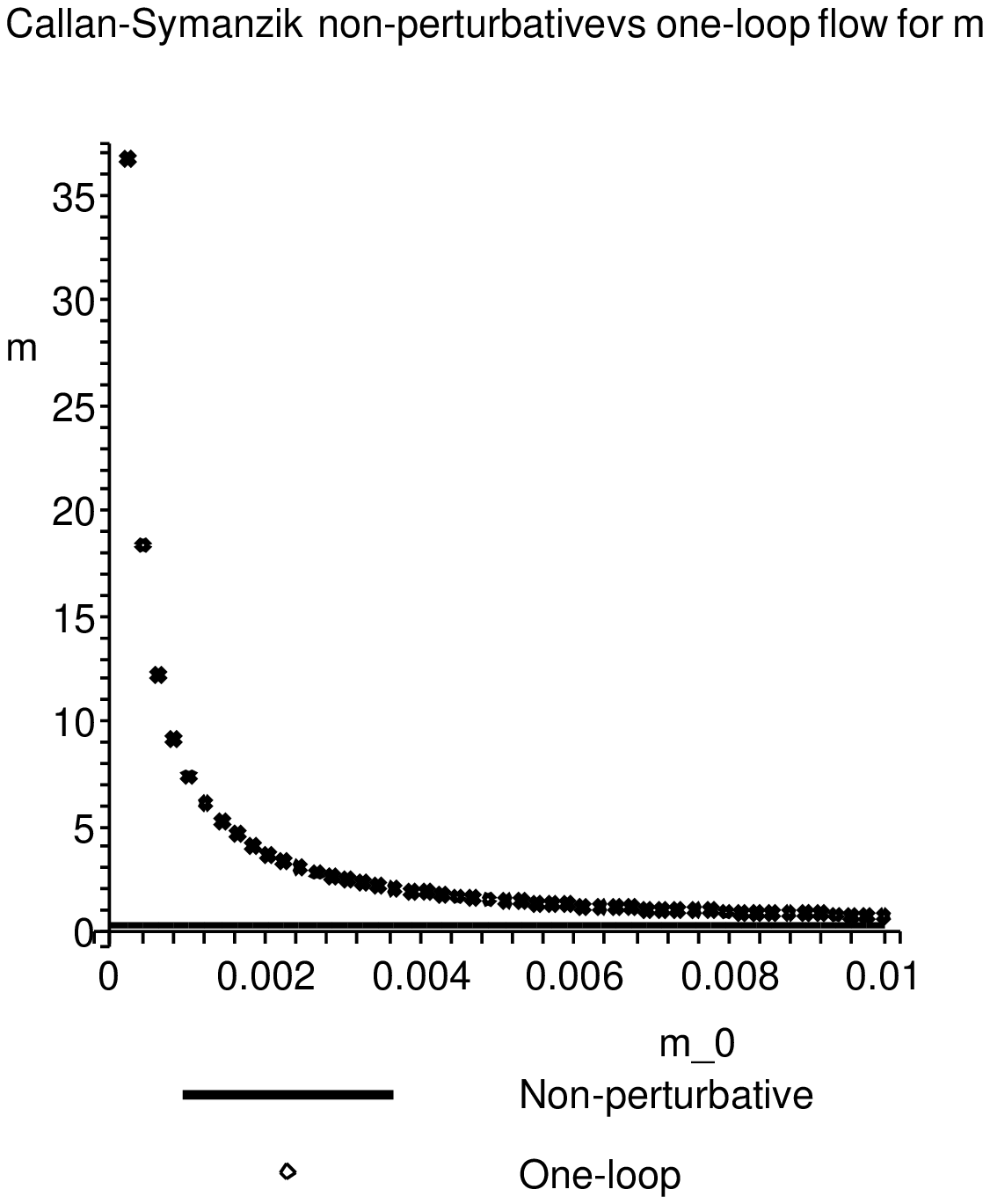}
\end{center}
\caption{Comparison between the one-loop and the non-perturbative flows for the fermion dressed mass
$m$  versus $m_0$
(the figure on the right is a zoom near $m_0=0$.}
\label{m1_comp}
\end{figure}

\section{Comments on the gauged model}

In this penultimate section, we would like to make some remarks concerning a possible extension of this Yukawa model
to a gauged one, in which the fermions couple to an Abelian Gauge field.
First, let us
discuss a few features of the Lifshitz-type $QED$, which may be relevant for our purposes.
Studies with anisotropic higher-order derivatives in $QED$
have already taken place in \cite{anselmi}, but there, the Lorentz-violating terms are added to the usual Lorentz invariant $QED$.
In the current article, we shall discuss a $QED$-type model, where {\it only} higher order derivatives are present in the model,
as is the case of the Lifshitz-Yukawa model studied in previous sections. \\
We first note that the action for an Abelian gauge field in this framework, reads:
\be\label{sgauge}
S_g=-\frac{1}{4}\int dtd^3{\bf x}\left\{2F_{0k}F^{0k}+G_{kl}G^{kl}\right\},
\ee
with $k,l=1,\cdots,D$ and
\be
F_{0k}=\partial_tA_k-\partial_kA_0~~~,~~~
G_{kl}=-\Delta(\partial_kA_l-\partial_lA_k),
\ee
where $\Delta=\delta^{kl}\partial_k\partial_l$. The gauge fields have dimensionalities
$[A_0]=2$ and $[A_k]=0$,
and the action (\ref{sgauge}) is invariant under the gauge transformation $A_\mu\to A_\mu+\partial_\mu\theta$,
where $[\theta]=-1$.\\
In order to find the gauge propagator, we first calculate
the second functional derivatives of the action (\ref{sgauge}) with respect to the gauge fields,
\bea
\frac{\delta^2 S_g}{\delta A_0\delta A_0}&=&-\Delta~\delta(t-t')\delta^{(D)}({\bf x}-{\bf x}')\\
\frac{\delta^2 S_g}{\delta A_0\delta A_k}&=&-\partial_t\partial^k~\delta(t-t')\delta^{(D)}({\bf x}-{\bf x}')\nn
\frac{\delta^2 S_g}{\delta A_k\delta A_l}&=&
\left[ \eta^{kl}(\partial_t^2-\Delta^3)-\Delta^2\partial^k\partial^l\right]
\delta(t-t')\delta^{(D)}({\bf x}-{\bf x}')~.\nonumber
\eea
In Fourier space, these read
\bea\label{second}
\frac{\delta^2 S_g}{\delta A_0\delta A_0}&=&{\bf p}^2~\delta(\omega-\omega')\delta^{(D)}({\bf p}-{\bf p}')\\
\frac{\delta^2 S_g}{\delta A_0\delta A_k}&=&\omega p^k~\delta(\omega-\omega')\delta^{(D)}({\bf p}-{\bf p}')\nn
\frac{\delta^2 S_g}{\delta A_k\delta A_l}&=&
\left[ (({\bf p}^2)^3-\omega^2)\eta^{kl}+({\bf p}^2)^2p^kp^l\right]
~\delta(\omega-\omega')\delta^{(D)}({\bf p}-{\bf p}')\nonumber.
\eea
Hence one can check that, for any $\mu=0,1,\cdots,D$,
\be
\frac{\delta^2S_g}{\delta A_\mu\delta A_\nu}~p_\nu=\frac{\delta^2S_g}{\delta A_\mu\delta A_0}~\omega
+\frac{\delta^2S_g}{\delta A_\nu\delta A_k}~p_k=0
\ee
which shows the absence of an inverse for the operator $\delta^2S_g$.

As usual, this is remedied by the addition of a gauge fixing term. In our Lifshitz case, this is naturally provided by the Coulomb gauge,
which is equivalent to adding the term
\be
-\frac{1}{2}\Delta^2\left(\partial^kA_k\right)^2
\ee
to the Lagrangian, so that the space components of the second derivatives
(\ref{second}) change to
\be
\frac{\delta^2 S_g}{\delta A_k\delta A_l}=(({\bf p}^2)^3-\omega^2)~\eta^{kl}
~\delta(\omega-\omega')\delta^{(D)}({\bf p}-{\bf p}').
\ee
Using the tensorial structure available in the $D$-dimensional space,
we search for a gauge field propagator $D_{\mu\nu}(\omega,{\bf p})$ in the form
\be
D_{00}=A,~~~~~~~~D_{0k}=Bp_k,~~~~~~~~D_{kl}=C\eta_{kl}+Ep_kp_l,
\ee
where $A,B,C,E$ are found from the definition
\be
iD_{\mu\rho}~\frac{\delta^2S_g}{\delta A_\rho\delta A_\nu}=
\delta^\nu_\mu~\delta(\omega-\omega')\delta^{(D)}({\bf p}-{\bf p}')~.
\ee
This leads to the following structures
\bea
D_{00}&=&\frac{-i}{{\bf p}^2}\left( 1-\frac{\omega^2}{({\bf p}^2)^3}\right) \\
D_{0k}&=&i\frac{\omega p_k}{({\bf p}^2)^4}\nn
D_{kl}&=&\frac{-i}{({\bf p}^2)^3-\omega^2}\left( \eta_{kl}+\frac{\omega^2p_kp_l}{({\bf p}^2)^4}\right)  \nonumber.
\eea
In the Yukawa model discussed above, the scalar field is real and has no $U(1)$ charge. However, the (Dirac) fermion
field can couple to the gauge field. This
can be done in several ways, respecting gauge invariance.
The minimal coupling can have one of the following forms
\bea\label{minimal}
&&\ol\psi(i\partial_t-eA_0)\gamma^0\psi-\ol\psi\left[\left(i\vec\partial-e {\bf A}\right)\cdot\vec\gamma\right]^3\psi \nn
&&\ol\psi(i\partial_t-eA_0)\gamma^0\psi-\ol\psi\left(i\vec\partial-e {\bf A}\right)^2
\left(i\vec\partial-e {\bf A}\right)\cdot\vec\gamma~\psi
\eea
Since the gauge coupling is dimensionful ($[e]=1$), the quantum theory will exhibit fermion dynamical mass generation
(the gauge boson remains massless because of gauge invariance).
Therefore the model cannot describe massless fermions and, as in the
Yukawa model, the fermion group velocity will be smaller than the speed of light. \\
The Lorentz-restoring kinetic terms arising from quantum fluctuations, for both
the fermion and the gauge field, contain several contributions because of the new vertices appearing in the
minimal couplings (\ref{minimal}). Nevertheless, at one loop, the only contribution to the space components of the
polarization tensor, which depends on the external momentum, is the same as the corresponding one in the Yukawa model, and is proportional to
\be
\frac{g_0^2}{m_0^2}({\bf p}^2-p^i p^j)~.
\ee
As we have already seen in the context of the Yukawa model above, this will lead to a speed of the gauge boson proportional 
to $g_0/m_0^3$.
If we expect this gauge boson to represent a physical photon, then we need this speed to be of the order of the
speed of light in vacuo (i.e. unity in our system of units). This implies that we have to be in a non-perturbative regime, where
$g_0$ is of order of $m_0^3$, and, thus, the one-loop analysis is no longer valid.
For this reason, a realistic study of the IR regime of this theory necessitates going
beyond the one-loop approximation, which lies outside the scope of the current article, and thus is left for a future work.

\section{Conclusions and Outlook}

In this work we have examined the infrared limit of a Lifshitz-type Lorentz-violating Yukawa model, in (3+1)-dimensional Minkowski
space time, where {\it only} higher order derivatives are present in the model. This differentiates the model, from existing ones
in the literature, e.g.~\cite{anselmi}, where the Lorentz-violating terms co-exist, as small corrections, with the standard
Lorentz-invariant ones. In our case, it is demonstrated that quantum corrections \emph{restore} relativistic-like kinematics in the low
energy-momentum regime (IR) of the model. 
In this way one can talk about a \emph{dynamically emergent} Lorentz Invariance in the model, where each particle sees a different 
effective light cone. The group velocities
of the fermions are found much smaller than those of the scalar field, although both velocities are much smaller compared to the
speed of light \emph{in vacuo}, consistent with the IR limit, where the analysis is performed.

Moreover, we have discussed a detailed mechanism for dynamical mass generation in the model in the same regime of low-energies.
It is demonstrated that there is no massless limit in this model, in the sense that the quantum-fluctuations dressed masses never
vanish, even in the limit where the bare masses go to zero.
The model exhibits infrared stability, in the sense that the zero-bare-mass limit corresponds to infrared-divergence-free resummed
quantum corrections. In our approach the resummation is provided by a differential Schwinger-Dyson type of approach, in which the
relevant dressed parameters of the model appear in a set of coupled differential functional equations. We solved numerically these
equations for the dressed mass and Yukawa coupling, and arrived at the aforementioned results on the IR stability and non-vanishing
dynamically generated masses.

We also discussed briefly an extension of the model, which involves coupling of the fermions to an Abelian gauge field.
The induced dispersion relations for the gauge field, which remains massless if we want to preserve the gauge invariance,
are such that its velocity is also found much smaller than the speed of light \emph{in vacuo}, thereby hampering any interpretation
of such gauge fields as the physical photons.

In view of the above properties, a natural question arises as to what physical systems, if any, such classes of models might
correspond to. In view of the \emph{emergent} Lorentz symmetry, one might view our model as a toy model for explaining a rather
\emph{microscopic origin of Lorentz symmetry}, in the low-energy limit, in analogy with the case of condensed matter systems,
with nodes in their Fermi surface, where linearization about them leads to low-energy relativistic excitations. Such a
Lifshitz-type scenario, with emergent Lorentz symmetry, could survive the full inclusion
of the Standard Model group, since the presence of several coupling constants would allow a fine tuning of parameters of the model, 
in such a way that the effective light cone seen by different species of particles could be made identical.

Another possible application of the (gauged) Yukawa model studied here, would be its association with the continuum limit of
some Lattice models of potential relevance to some condensed matter systems in four space-time dimensions, in analogy with the
case of planar high temperature superconductivity, where the gauge fields represent fractional statistics in (2+1)-dimensions.
In such models, the gauge field is also different from the real photon, as in the gauged model of section 4. Thus, one may
view the neutral scalar as a (spinless) phonon excitation, e.g. due to lattice ion vibrations, while the charged Dirac fermion could represent a spin mode coupled to an appropriate gauge potential, which, however, is not the physical photon.
In condensed matter systems, the relativistic fermions may be associated with excitations near a node in the Fermi surface of the microscopic model~\cite{hightc}.
From our analysis in section 4, one may assume that the massless gauge boson group velocity may represent the velocity of
the node of the Fermi surface about which we linearized the continuum-model. In such a case, the physical significance of the
gauge potential may be similar to that of a (3+1)-dimensional statistics-changing field for the fermion excitations, which could
represent some kind of hole excitations, due to doping.

\vspace{1cm}

\nin{\bf Acknowledgements} J.~A. would like to thank P. Pasipoularides and A. Tsapalis for useful comments.
This work is partially supported by the Royal Society, UK, the European Union (N.E.M.),
through the Marie Curie Research and
Training Network \emph{UniverseNet} (MRTN-2006-035863), and the STFC UK (D.~Y.), through a Graduate Research Quota
Studentship.

\section*{Appendix A: One loop effective theory}

\subsection*{One-loop scalar self energy}

For external momentum ${\bf k}$ and vanishing external frequency, the one-loop scalar self energy is, after a Wick rotation,
\bea\label{sigmas}
\Sigma_s&=&-ig_0^2~\mbox{tr}\int\frac{d\omega}{2\pi}\frac{d^3{\bf p}}{(2\pi)^3}
\frac{i\omega\gamma^0-p^2({\bf p}\cdot\vec\gamma)+m_0^3}{\omega^2+p^6+m_0^6}\nn
&&~~~~~\times\frac{i\omega\gamma^0-({\bf p}+{\bf k})^2(({\bf p}+{\bf k})\cdot\vec\gamma)+m_0^3}{\omega^2+({\bf p}+{\bf k})^6+m_0^6}.
\eea
The scalar mass correction is given by the zeroth order in ${\bf k}$, which is
\bea
\Sigma_s^{(0)}&=&4ig_0^2\int\frac{d\omega}{2\pi}\frac{d^3{\bf p}}{(2\pi)^3}
\frac{\omega^2+p^6-m_0^6}{(\omega^2+p^6+m_0^6)^2}\nn
&=&\frac{ig_0^2}{\pi^2}\int\frac{p^8dp}{(p^6+m_0^6)^{3/2}}\nn
&=&\frac{ig_0^2}{\pi^2}\left[ \ln\left( \frac{\Lambda}{m_0}\right) +\frac{\ln2-1}{3}\right] +{\cal O}(\Lambda^{-2}),
\eea
and we finally obtain for the scalar mass
\be
M^6_{(1)}=m_0^6+\frac{g_0^2}{\pi^2}\left[ \ln\left( \frac{\Lambda}{m_0}\right) +\frac{\ln2-1}{3}\right]~.
\ee
The term quadratic in spatial derivatives is given by the quadratic-order term in the Taylor expansion of the self energy (\ref{sigmas})
in powers of ${\bf k}$:
\bea
\Sigma_s^{(2)}&=&4ig_0^2\int\frac{d\omega}{2\pi}\frac{d^3{\bf p}}{(2\pi)^3}\Bigg\{
\frac{p^4k^2+2p^2({\bf p}\cdot{\bf k})^2}{{\cal D}^2}-18p^8\frac{({\bf p}\cdot{\bf k})^2}{{\cal D}^3}\\
&&+\frac{\omega^2+p^6-m_0^6}{{\cal D}^3}
\left( -3p^4k^2-12p^2({\bf p}\cdot{\bf k})^2+\frac{36p^8({\bf p}\cdot{\bf k})^2}{{\cal D}}\right)\Bigg\}\nonumber
\eea
where ${\cal D}=\omega^2+p^6+m_0^6$. Using then the following identity, valid for any function $f$,
\be
\int\frac{d^3{\bf p}}{(2\pi)^3}({\bf p}\cdot{\bf k})^2f(p^2)=\frac{4\pi k^2}{3(2\pi)^2}\int dp ~p^4f(p^2),
\ee
we obtain
\bea
\Sigma_s^{(2)}&=&\frac{ig_0^2 k^2}{\pi^3}\int dp~p^2\int d\omega\left( -\frac{16p^4}{3{\cal D}^2}+\frac{6p^{10}+14m_0^6p^4}{{\cal D}^3}
-\frac{24m^6p^{10}}{{\cal D}^4}\right) \nn
&=&-\frac{ig_0^2k^2}{\pi^2m_0^2}\int_0^\infty \frac{dx~x^6}{(1+x^6)^{3/2}}\left(\frac{8}{3}-\frac{3(3x^6+7)}{4(1+x^6)}+\frac{15x^6}{2(1+x^6)^2}\right).
\eea
The latter result has to be identified with $i\lambda_\phi^{(1)} k^2$,
which leads to the expression (\ref{lambdasoneloop}) for $\lambda_\phi^{(1)}$, with a negative sign.

\subsection*{One-loop fermion self energy}

The one-loop fermion self energy is, for external momentum ${\bf k}$ and vanishing external frequency, after a Wick rotation,
\be
\Sigma_f=ig^2\int \frac{d^3{\bf p}}{(2\pi)^3}\frac{d\omega}{2\pi}
\frac{\omega\gamma^0-p^2({\bf p}\cdot\vec\gamma)+m_0^3}{{\cal D}}
\frac{1}{\omega^2+({\bf p}-{\bf k})^6+m_0^6}.
\ee
The correction to the fermion mass is obtained from the zeroth-order term in ${\bf k}$:
\bea
\Sigma_f^{(0)}&=&\frac{ig_0^2}{(2\pi)^4}\int d^3{\bf p}d\omega\frac{m_0^3}{{\cal D}^2}\nn
&=&\frac{ig_0^2m_0^3}{8\pi^2}\int \frac{p^2dp}{(p^6+m_0^6)^{3/2}},
\eea
which yields
\be
m^3=m_0^3+\frac{g_0^2}{24\pi^2m_0^3}.
\ee
The term linear in the spatial derivatives is given by the first-order term in the Taylor expansion in powers of ${\bf k}$:
\bea
\Sigma_f^{(1)}&=&-\frac{6g_0^2}{(2\pi)^4}\int d^3{\bf p}d\omega\frac{p^6({\bf p}\cdot\vec\gamma)({\bf p}\cdot{\bf k})}{{\cal D}^3}\nn
&=&-\frac{ig_0^2}{2\pi^3}({\bf k}\cdot\vec\gamma)\int \frac{p^{10}dpd\omega}{{\cal D}^3}\nn
&=&\frac{-3ig_0^2}{16\pi^2}({\bf k}\cdot\vec\gamma)\int\frac{p^{10}~dp}{(p^6+m_0^6)^{5/2}}.
\eea
This expression has to be identified with $-i\lambda_\psi^{(1)} ({\bf k}\cdot\vec\gamma)$, which leads to the expression
\be
\lambda_\psi^{(1)}=\frac{3g_0^2}{16\pi^2}\int \frac{p^{10}~dp}{(p^6+m_0^6)^{5/2}}.
\ee

\subsection*{One-loop vertex}

We show here that the one-loop correction to the coupling vanishes.
This correction is given by the following three-point graph, for vanishing incoming momentum:
\be
g^{(1)}=g_0+ig_0^3~\mbox{tr}\int\frac{d\omega}{2\pi}\frac{d^3{\bf p}}{(2\pi)^3}
\frac{\left( \omega\gamma^0-{\bf p}^2 ({\bf p}\cdot\gamma)+m_0^3\right) ^2}{\left( \omega^2-p^6-m_0^6+i\varepsilon\right)^3},
\ee
and, after a Wick rotation,
\bea
g^{(1)}
&=&g_0+\frac{g_0^3}{\pi^3}\int d\omega ~p^2dp\frac{-\omega^2-p^6+m_0^6}{(\omega^2+p^6+m_0^6)^3}\nn
&=&g_0+\frac{g_0^3}{4\pi^2}\int p^2dp\left(\frac{3m_0^6}{(p^6+m_0^6)^{5/2}}-\frac{2}{(p^6+m_0^6)^{3/2}}\right)\nn
&=&g_0+\frac{g_0^3}{12\pi^2m_0^6}\int_0^\infty dx\left( \frac{3}{(1+x^2)^{5/2}}-\frac{2}{(1+x^2)^{3/2}}\right)\nonumber
\eea
The last integral vanishes, and $g^{(1)}=g_0$.

\section*{Appendix B: Evolution equations for the scalar potentials}

Taking into account the gradient expansion (\ref{ansatz}) for the constant-field case
$\phi_{cl}\rightarrow\phi_0$\ and $\ol\psi_{cl}\psi_{cl}\rightarrow \xi_0$ leads to
\bea\label{phicompact}
\frac{\delta^2\Gamma}{\delta\phi_p\delta\phi_q}&=&\bigg[\omega^2-p^6
-U''\xi_0 -V''\bigg]\delta^{4}(p+q)
\equiv\Gamma^{(2)}_{\phi \phi}\delta^{4}(p+q)\nn
\frac{\delta^2\Gamma}{\delta\ol\psi_p\delta\psi_q}&=&
\bigg[\gamma^0\omega-p^2({\bf p}\cdot\vec\gamma)+U\bigg]\delta^{4}(p+q)
\equiv\Gamma^{(2)}_{\ol\psi \psi}\delta^{4}(p+q)\nn
\frac{\delta^2\Gamma}{\delta\psi\delta\ol\psi}&=&\bigg[\gamma^0\omega-p^2({\bf p}\cdot\vec\gamma)
-U\bigg]\delta^{4}(p+q)\equiv\Gamma^{(2)}_{\psi \ol\psi}\delta^{4}(p+q)\nn
\frac{\delta^2\Gamma}{\delta\phi_p\delta\ol\psi_q}&=&-U'\psi_{cl}\delta^4(p+q)
\equiv\Gamma^{(2)}_{\ol\psi \phi}\delta^4(p+q)\nn
\frac{\delta^2\Gamma}{\delta\phi_p\delta\psi_q}&=&U'\ol\psi_{cl}\delta^4(p+q)
\equiv\Gamma^{(2)}_{\psi_p\phi_q}\delta^4(p+q)
\eea
We have to calculate the inverse of the matrix
\be
\Gamma^{(2)}=\left(\begin{array}{ccc}
\Gamma^{(2)}_{\ol\psi \psi} & \Gamma^{(2)}_{\ol\psi \ol\psi} & \Gamma^{(2)}_{\ol\psi \phi} \\
\Gamma^{(2)}_{\psi \psi} & \Gamma^{(2)}_{\psi\ol\psi} & \Gamma^{(2)}_{\psi \phi}\\
\Gamma^{(2)}_{\phi \psi} & \Gamma^{(2)}_{\phi\ol\psi} & \Gamma^{(2)}_{\phi \phi} \\
\end{array}\right),
\ee
that we find by using the following expansion, valid for any two $3\times3$ matrices $A,B$
\be
(A+\xi_0B)^{-1}=A^{-1}-A^{-1}BA^{-1}\xi_0+{\cal O}(\xi_0^2).
\ee
In our case, this expansion gives for the elements of the inverse matrix $[\Gamma^{(2)}]^{-1}$,
\bea\label{expansion}
&&[\Gamma^{(2)}]^{-1}_{\phi\phi}=\frac{1}{\omega^2-p^6-V''}\\
&&+\frac{\xi_0}{(\omega^2-p^6 -V'')^2}\times
\left[U''-\frac{2[U']^2U}{(\gamma^0 \omega-p^2({\bf p}\cdot\vec\gamma))^2-U^2}\right]+{\cal O}(\xi_0^2),\nonumber\\
&&=I_1+\xi_0 I_2+{\cal O}(\xi_0^2),\nonumber
\eea
and
\bea
&&[\Gamma^{(2)}]^{-1}_{\ol\psi\psi}
=\frac{1}{\gamma^0 \omega-p^2({\bf p}\cdot\vec\gamma)+ U}\\
&&+\frac{-2\xi_0[U']^{2}U}{\bigg[\gamma^0 \omega-p^{2}({\bf p}\cdot\vec\gamma)
+ U\bigg]\bigg[\left(\gamma^0 \omega-p^2({\bf p}\cdot\vec\gamma)\right)^{2}-U^{2}\bigg]\bigg[\omega^2-p^6-V''\bigg]}\nn
&&+\frac{-\xi_0[U']^{2}}{\bigg[(\gamma^0 \omega
-p^{2}({\bf p}\cdot\vec\gamma))^{2}-U^2\bigg]\bigg[\omega^2-p^6-V''\bigg]}+\mathcal{O}(\xi_{0}^{2})\nonumber\\
&&=I_3+\xi_0 I_4+{\cal O}(\xi_0^2),\nonumber
\eea
After a Wick rotation, the integration over $(\omega,{\bf p})$ gives,
for large values of $\Lambda$:
\bea
\mbox{Tr} \{I_1\}&=&\frac{-i{\cal V}}{12\pi^2}\ln\left(\frac{2\Lambda^3}{[V'']^\hf}\right)\\
\mbox{Tr}\{I_2\}&=&\frac{i{\cal V}}{24\pi^2}\frac{U''}{V''}+\frac{i[U']^{2}U}{6\pi^2}
\left[-\ln\left(\frac{U}{[V'']^\hf}\right)\frac{1}{(V''-U^2)^2}-\frac{1}{2(V''-U^2)V''}\right]\nn
\mbox{Tr}\{I_{3}\}&=&{\cal V}\frac{iU}{12\pi^2}\ln\left(\frac{2\Lambda^3}{U}\right)\nn
\mbox{Tr}\{I_{4}\}&=&{\cal V}\frac{-i[U']^{2}U^2}{6\pi^2}\left[\ln\left(\frac{U}{[V'']^\hf}\right)
\frac{1}{(V''-U^2)^2}+\frac{1}{2(V''-U^2)U^2}\right]\nn
&+&\frac{i{\cal V}}{12\pi^2}\frac{[U']^2}{(V''-U^2)}\ln\left(\frac{U}{[V'']^\hf}\right)\nonumber,
\eea
where ${\cal V}$ is the space time volume. The reader should notice that this volume factor cancels out in eq.(\ref{evolG}),  since it appears on both sides of the equation.

\end{document}